\documentclass{emulateapj}

\newcommand{\kms}{\ifmmode\,{\rm km}\,{\rm s}^{-1}\else km$\,$s$^{-1}$\fi}
\newcommand{\Rd}{\ifmmode\,R_{\rm d}\else $R_{\rm d}$\fi}
\newcommand{\be}{\begin{equation}}
\newcommand{\ee}{\end{equation}}
\newcommand\ltsima{$\; \buildrel < \over \sim \;$}
\newcommand\ltsim{\lower.5ex\hbox{\ltsima}}
\newcommand\gtsima{$\; \buildrel > \over \sim \;$}
\newcommand\gtsim{\lower.5ex\hbox{\gtsima}}
\def\Eq#1{Eq.~(\ref{eq:#1})}

\def\Fig#1{Fig.~\ref{fig:#1}}

\slugcomment{Accepted for publication in Astrophysical Journal Letters}
\begin{document}

\title{On the Global Mass Distribution in Disk Galaxies}

\author{
St\'ephane Courteau\altaffilmark{1}
\& Aaron A. Dutton\altaffilmark{2}
}
\altaffiltext{1}{Department for Physics, Engineering Physics and Astrophysics, Queen's University, Kingston, ON K7L 3N6, Canada }
\altaffiltext{2}{Max-Planck-Institut f\"{u}r Astronomie, K\"{o}nigstuhl 17, D-69117 Heidelberg, Germany}

\begin{abstract}
  We present a summary of recent assessments of the mass distribution
  in disk galaxies.  Of issue in order to characterise
  galaxy formation models is to determine the relative fraction of
  baryons and dark matter at all radii in galaxies. For disk galaxies,
  various measurements of the mass distribution in galaxies based on
  vertical kinematics, strong lensing, residuals of scaling relations,
  fluid dynamical modeling, bar strength and pattern speed, warps,
  and others, have called for either a maximal or sub-maximal
  contribution of the baryons in the inner parts of the disk.
  We propose a global picture whereby all galaxies are typically
  baryon-dominated (maximal) at the center and dark-matter dominated
  (sub-maximal) in their outskirts.  Using as a fiducial radius
  the peak of the rotation curve of a pure baryonic exponential
  disk at $R_{2.2}=2.2\Rd$, where $\Rd$ is the disk scale length,
  the transition from maximal to sub-maximal baryons occurs within
  or near $R_{2.2}$ for low-mass disk galaxies
  (with $V_{\rm tot} \ltsim 200$ \kms) and beyond $R_{2.2}$
  for more massive systems.  The mass fraction's inverse dependence
  on circular velocity in disk galaxies is largely insensitive to
  the presence of a bar at $R_{2.2}$.  The mean mass fractions of
  late- and early-type galaxies are shown to be different at the same
  fiducial radius and circular velocity, pointing to different galaxy
  formation mechanisms.  Feedback, dynamical friction, size, and IMF
  variations are key ingredients for understanding these differences.
\end{abstract}


\section{Introduction}

Ever since the first speculations about the presence of ``dark
matter'' in galaxies \citep{Kapteyn22,Oort32}, astronomers have
endeavored to characterize the fraction of visible to total mass at
all radii in galaxies. The relative distribution of matter in galaxies
ought to be one of the most definitive predictions of galaxy formation
models yet its validation is challenged by numerous observational,
theoretical, and operational challenges.  While galaxies are believed
to be dominated by an invisible matter component in their outskirts, a
debate has been blazing for the last two decades regarding the
relative fraction of baryons and dark matter in the inner parts of
galaxies: whether galaxies are centrally dominated by baryons
(``maximal disk'') is of issue.  Some of those debates have been
misconstrued on account of operational confusion, such as dark matter
fractions being measured and compared at different radii.  An ultimate
goal for galaxy structure studies is to achieve {\it accurate}
data-model comparisons for the relative fractions of baryonic to total
matter at any radius.

In this Letter, we wish to provide a global appreciation of the mass
distribution of baryons and dark matter in disk galaxies by combining
various reports of the baryon velocity fraction, ${\cal{F}}(R) \equiv
V_{\rm \rm bar}/V_{\rm tot}$ in disk galaxies, where $V_{\rm bar}$ is
the inferred circular velocity ascribed to the baryonic material and
$V_{\rm tot}$ is the total measured circular velocity at a given
radius.  $V_{\rm bar}$ is typically inferred by multiplying a galaxy
light profile, $L(R)$, by a suitable mass-to-light ratio,
$\left(M_*/L\right) (R)$.  Any difference between $V_{\rm bar}$ and
$V_{\rm tot}$ is viewed as a signature for the presence of dark
matter. The dark matter fraction is defined as 
$f_{\rm DM}(R)= M_{\rm DM}(R)/M_{\rm tot}(R) \approx 1 - {\cal{F}}^2$,
where $M_{\rm DM}(R)$ and $M_{\rm tot}(R)$ are the dark matter and total
mass enclosed within radius $R$, respectively. The approximation
becomes an equality for spherical systems.

How galaxies ultimately arrange their baryonic and non-baryonic matter
is the result of numerous complex mechanisms involving their mass
accretion history, the depth of the potential well, the initial mass
function (which ultimately affects feedback and quenching processes),
dynamical friction, dynamical instabilities, and more.  Analytical
models of galaxy formation strive to predict $f_{\rm DM}(R)$
\citep{Dalcanton97,Dutton07,Mo10} with mitigated success in light of
the above challenges; numerical simulations suffer additional
limitations too.

Can observations provide unique determinations of $f_{\rm DM}(R)$ as
ideal constraints to galaxy formation models?  Not yet, but, after
decades of muddled debates and incomplete data, a clear picture seems
to be emerging.

\section{Mass Fraction Assessment}

Mass models of disk galaxies, where the observed rotation curve is
decomposed into its principal gas, stars, and dark matter components,
have historically been embraced as the ideal method to separate
baryons and dark matter as a function of radius
\citep{Bosma78,Carignan85,vanAlbada85}.  The pros and cons of this
approach are addressed in the review on ``Galaxy Masses'' by
\cite{Courteau14}.  As stressed in that review \citep[see
also][]{Dutton05}, fundamental degeneracies between the disk and dark
halo models prevent a unique baryon/dark matter decomposition based on
rotation curve data alone.  Other information or methods must be
considered to break these degeneracies.

Among others, the quantity $V_{\rm bar}$ is obtained indirectly via
a stellar or baryonic mass-to-light ratio ($M_*/L$) which is itself
inferred from stellar population studies or dynamical stability
analysis.  Current stellar population mass-to-light ratios carry an
uncertainty of a factor 2 at best \citep{Conroy13,Courteau14}.  The
latter uncertainty is unfortunately large enough to encompass a full
range of baryon fractions at a given radius.
  
For reference, a maximal disk is often defined to obey 
\be
{\cal{F}} \equiv V_{\rm disk}(R_{\rm max})/V_{\rm tot}(R_{\rm max}) >
0.85,
\label{eq:maxdisk}
\ee where $V_{\rm disk}$ is the inferred velocity of the disk (stars
and gas), $V_{\rm tot}$ is the total observed velocity, and 
$R_{\rm max}$ is the radius at which $V_{\rm disk}$ reaches its peak value
\citep{vanAlbada85,Sackett97}.

In terms of dark matter fraction this criterion corresponds to
$f_{\rm DM} < 0.28$.

\subsection{Tully-Fisher residuals}
To circumvent the loopholes of mass modeling, \cite{CourteauRix99}
observed that the residuals of the velocity-luminosity relation,
$\Delta \log V(L)$, contrasted against those of the size-luminosity
relation, $\Delta \log R(L)$, can provide a tighter constraints on
$\cal{F}$ measured at $2.2\Rd$.  

It can be easily shown that $\Delta \log V(L)/\Delta \log R(L) = -1/2$
for baryon dominated galaxies \citep{CourteauRix99} and $\Delta \log
V(L)/\Delta \log R(L) > 0$ for dark matter dominated galaxies.
Numerous scaling relation analyses of spiral galaxies have found that
$\Delta \log V(L)/\Delta \log R(L) \simeq 0$
\citep{CourteauRix99,Courteau07,Dutton07,Reyes11}.  The null slope
indicates that both baryons and dark matter contribute to the galaxy
dynamics at $2.2 \Rd$.  Various galaxy structure models whereby the
dark matter is compressed adiabatically by the cooling baryons as the
galaxy stabilizes dynamically yield a mapping between $\Delta \log
V(L)/\Delta \log R(L)$ and the baryon fraction $\cal{F}$ (see
references above).  These models yield, on average for bright spiral
galaxies, ${\cal{F}}(2.2\Rd)=0.55-0.71$.  The lower value applies if
only adiabatic contraction is at play; a higher value is found if
other mechanisms (e.g. feedback) oppose the adiabatic contraction
\citep[]{CourteauRix99,Dutton07}.  By virtue of \Eq{maxdisk} above,
typical galaxy disks are thus, {\it on average}, sub-maximal at
$2.2\Rd$.  Since mass scale as $V^2$, the dark matter fraction,
$f_{\rm DM}(2.2\Rd)\simeq 50-70\%$.  \cite{Reyes11} suggested that
$\cal{F}$ is controlled by the surface density of the baryonic
material; i.e. if the latter is high, so is $\cal{F}$ \citep[see also
Fig. 23 in][]{Bovy13}.

\subsection{Vertical Disk Kinematics}
A similar result is obtained via the ``velocity dispersion'' method of
\citep{vdK88}, as applied by many for a full suite of disk galaxies
\citep{Bottema93}, edge-on systems \citep{Kregel05}, or face-on spiral
galaxies \citep{Bershady11,Martinsson13}. For a self-gravitating,
radially exponential disk with vertical profile of the form
$\rho(R,z)=\rho(R,0) {\rm sech}^2(z/z_0)$, \cite{vdK88} and
\cite{Bottema93} showed that the peak circular velocity of the stellar
disk, $V_{\rm{disk}}$, measured at $R=2.2\Rd$ can be related to
the vertical velocity dispersion, $V_z$, and the intrinsic thickness
(or scale height) of the disk, $z_0$, via:
\be
V_{\rm disk}(R_{\rm peak}) = c \, {\left\langle V_z^2 \right\rangle}^{1/2}_{R=0}
\,\sqrt{\frac{R_{\rm d}}{z_0}}.
\label{eq:Freeman}
\ee
where $c \simeq 0.88 (1-0.28 z_0/R_{\rm d})$ \citep{Bershady11}.

$V_{\rm disk}(R_{\rm peak})$, the contribution of the baryonic disk to
the total velocity where the rotation curve reaches its peak (at
$R_{\rm peak}$), can thus be measured directly {\bf\underbar{if}} the
disk scale length, $R_{\rm d}$, the scale height, $z_0$, and the
vertical component of the velocity dispersion, $V_z$, are known.  For
face-on systems, $V_z$ and $R_{\rm d}$ can be measured but $z_0$ must be
inferred; and vice versa for edge-ons.  Van der Kruit's method is thus
statistical in nature and relies on a number of (potentially noisy)
scaling transformations.  \cite{Bovy13} also point out that the
velocity dispersions measured from integrated light, as in most
applications of \Eq{Freeman} for external galaxies so far, do not
trace the older, dynamically-relaxed stellar populations with high
fidelity. The younger, brighter, stellar populations traced by the
light-weighted spectra tend to have a lower velocity dispersion
therefore biasing $V_{\rm disk}(R_{\rm peak})$ low and thus favoring
sub-maximal disks.  Sample selection also plays a substantial role
(see Fig.~1 below).

The DiskMass project has made use of van der Kruit's formalism to
extract $V_{\rm disk}(R_{\rm peak})$ for a sample of 46 nearly face-on
(inclinations $\simeq 30$ degrees) galaxies with rotation velocities
between 100 \kms\ and 250 \kms \citep{Bershady11}. This survey uses
integral-field spectroscopy to measure stellar and gas kinematics
using the custom-built SparsePak and PPAK instruments.  For 30
DiskMass galaxies covering a range of structural properties,
\cite{Martinsson13} report that the fraction $\cal{F}$ ranges from
0.34 to 0.73 and increases with luminosity, surface brightness,
rotation speed, and redder color.  The average value of their sample
is ${\cal{F}} =0.57 \pm 0.07$, in agreement with \cite{CourteauRix99}.

\subsection{Other methods}
Similar and broader results have been achieved through very different
techniques for disk dominated systems: such as fluid dynamical
modeling \citep[][see the latter reference for a more critical
outlook]{Weiner01,Kranz03,Athanassoula14} and strong gravitational
lensing \citep{Barnabe12,Dutton13a}.  For instance, because
gravitational lensing is sensitive to projected mass while dynamics
trace the mass within ellipsoids, their combination provides
additional geometrical dimensions to disentangle the baryons and dark
matter in galaxies; this approach is especially efficient for
highly-inclined disk galaxies \citep{Dutton11b}.  These methods,
reviewed in \cite{Courteau14}, are consistent {\it on average} with
${\cal{F}}(2.2\Rd)=0.6-0.7$, but also show a broad range of
${\cal{F}}$ as a function of maximum circular velocity or disk size.
In their hydrodynamical modeling of five grand design non-barred
galaxies, \cite{Kranz03} presciently suggested that galaxy disks
appear to be maximal if $V_{\rm max} > 200$ \kms, sub-maximal
otherwise.  The later summarizes rather well the more complete picture
about the mass distribution in galaxies that is now clearly emerging.

\section{A unified view}
\begin{figure}
\centering
\includegraphics[width=0.48\textwidth]{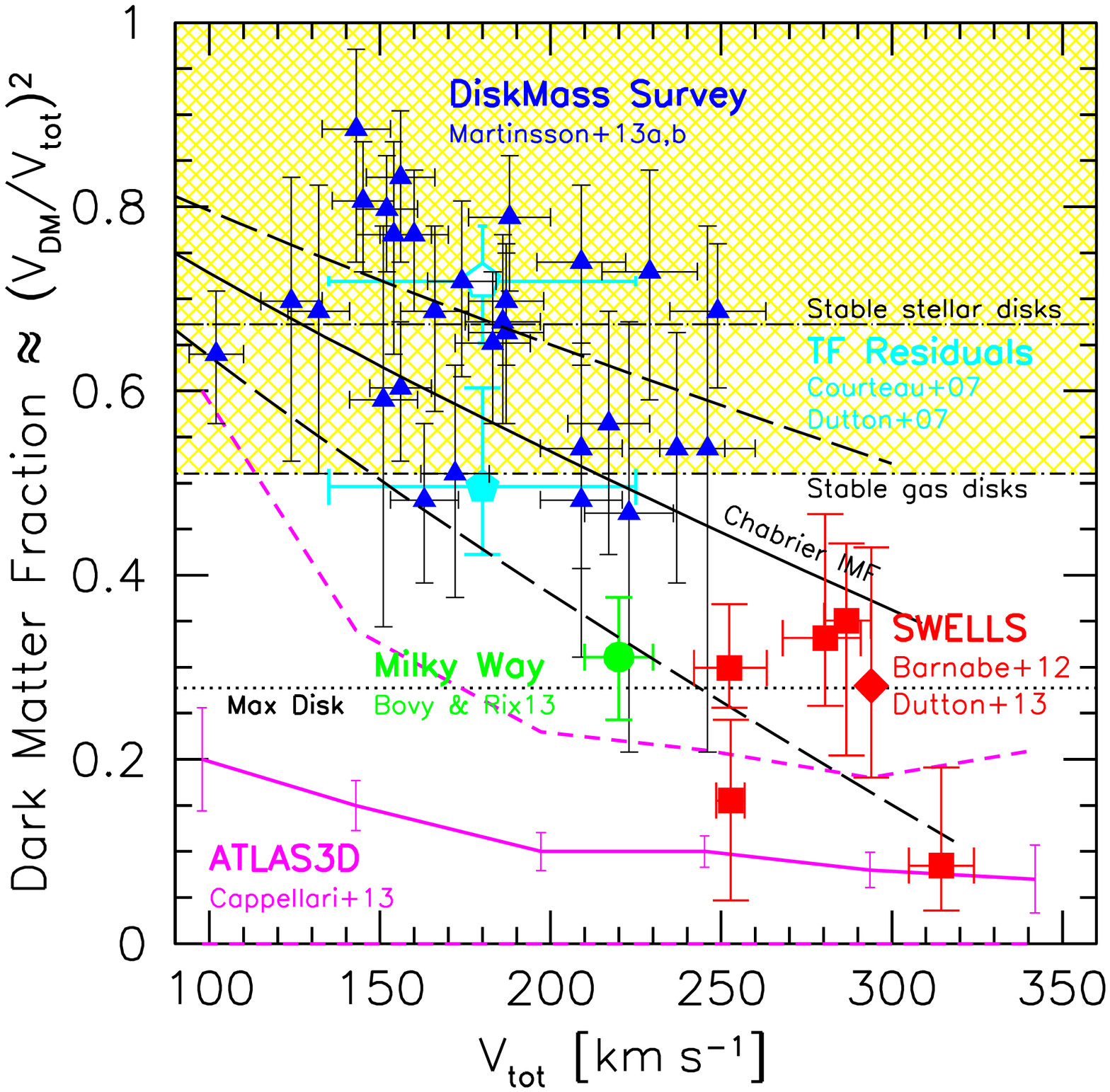}
\caption{Dark matter fraction at $2.2 \Rd$ as a function of total
  circular velocity for samples discussed in the text.  A clear trend
  is observed such that the dark matter fraction decreases with
  circular speed; indeed below 250 \kms, dark matter dominates
  ($>50\%$) the total mass budget of the galaxy at 2.2 $\Rd$.  The
  yellow shading shows the stability criteria \citep{Efstathiou82} for
  stellar disks ($f_{\rm DM}\simeq 0.67$; short-dashed black line and
  above) and gaseous disks ($f_{\rm DM} \simeq 0.51$).  The dotted
  black line defines a maximal disk (Eq. [1]).  The black line represents
  a model for galaxy scaling relations assuming a Chabrier initial mass
  function \citep{Dutton11a}; the long-dashed black lines are the 1$\sigma$
  dispersion in galaxy sizes.}
\label{fig:figDM}
\end{figure}

\Fig{figDM} is a compilation of dark matter fractions, $f_{\rm DM} = 1 -
{\cal{F}}^2 \approx (V_{\rm DM}/V_{\rm tot})^2$, at $\sim 2.2 \Rd$ for various
galaxy samples discussed above (others are discussed below), as a
function of total circular velocity.

The DiskMass survey (blue points) predominantly samples
large-to-average size galaxies thus avoiding the regime of disk
maximality (at $2.2 \Rd$).  The cyan pentagons represent the parameter
space for the Tully-Fisher residual analysis of the \cite{Courteau07}
sample of spiral galaxies with (open pentagon) and without (filled
pentagon) adiabatic contraction \citep{Dutton07}.  The Courteau and
DiskMass samples target essentially the same sub-maximal galaxy types.
Unlike those samples, the SWELLS gravitational lensing survey is
intrinsically biased towards high mass systems
\citep{Barnabe12,Dutton13a}.  The Milky Way study of \cite{Bovy13}
uses a method similar to that of \cite{vdK88} but with the added
benefit that stellar populations can be disentangled, thus providing
less biased measurements of the true underlying structure (scale
length, rotation and dispersion) of the dynamical tracer.  Their
analysis results in a low mass-weighted disk scale length, $\Rd = 2.15
\pm 0.14$ kpc, and a correspondingly low estimate of $f_{\rm
  DM}=0.31\pm0.07$ at 2.2$\Rd$\footnote{Various investigations of the
  Milky Way mass density profiles are reported in Table 1 of
  \cite{Courteau14}.  The early studies of \cite{DehnenBinney98} and
  \cite{Englmaier99} already pointed to a maximal MW disk.
  Measurements of $\cal{F}$ for the Milky Way hinge significantly on
  estimates of its disk scale length.  A low estimate of $\Rd$ yields
  a high value of $\cal{F}$ (or $f_{\rm DM}$).}.

The short-dash boundary delineates the region where baryon-heavy disk
galaxies become disk unstable.  The criterion for stability against a
bar is $V_{\rm circ} (\Rd / G M_{\rm disk})^{1/2} \gtsim$ 1.1
\citep{Efstathiou82,Foyle08}.  The yellow shading shows the stability
criteria for stellar ($f_{\rm DM} > 0.67$) and gaseous ($f_{\rm DM} >
0.51$) disks.  Barred and unbarred galaxies have also been shown to
have the same velocity-luminosity and size-luminosity scaling
relations (measured at $R_{2.2}$) \citep{Courteau03,Sheth12}.  Thus,
in a global dynamical sense, barred and unbarred galaxies behave
similarly and have, on average, comparable fractions of luminous and
dark matter at a given radius.

The black lines in \Fig{figDM} show a toy model where we have used the
observed scaling relations \citep{Courteau07,Dutton11a} between disk
scale length, rotation velocity and stellar mass (assuming a Chabrier
IMF) to estimate the dark matter fraction, assuming the baryonic mass
is dominated by an exponential stellar disk (i.e., gas and stellar
bulges are ignored).  The dashed lines show the effect of changing
the disk sizes by $\pm 0.16$ dex (the scatter in the size-velocity
relation). Changing the stellar masses (e.g., by changing the IMF)
by the same amounts leads to similar changes in dark matter
fractions. This simple model captures the main features of the
observations: 
\begin{enumerate}
\item Dark matter fraction decreases with galaxy circular velocity;
\item Milky Way mass galaxies ($V_{\rm circ}\sim 220$ \kms)
  have, on average at $2.2 \Rd$, $f_{\rm DM}\sim 0.5$; and
\item There is significant scatter in dark matter fractions at a given
  circular velocity driven, at least partly, by size variations.
\end{enumerate}

\subsection{Early-Type Galaxies}
\label{sec:ETG}
Measuring dark matter fractions in early-type galaxies is more
challenging. Firstly, because measuring total mass profiles is subject
to the mass-anisotropy degeneracy, which requires high-quality data to
break. Secondly, because there are no direct dynamical means of
separating the total mass into baryons and dark matter. Observations
of dark matter fractions in early-type galaxies are thus driven by
systematics in the mass modeling assumptions, such as dark matter
profile, and stellar M/L gradients.

\Fig{figDM} also shows the distribution $f_{\rm DM}(R_{\rm e})$ of 260
ATLAS3D early-type galaxies \citep{ATLAS3D}, where we have used
$V_{\rm circ}=1.38 \sigma_8$ to transform velocity dispersions,
$\sigma_8$, to circular velocity $V_{\rm circ}$ \citep{Dutton11b}. An
important caveat to these results is the assumption of an NFW dark
matter halo (Navarro, Frenk, \& White 1997) in the mass modelling.
Dark matter fractions as high as $f_{\rm DM}(R_{\rm e})\sim 60\%$ are
obtained if dark matter haloes contract adiabatically in response to
galaxy formation \citep{Dutton13b}.  Note also that ATLAS3D velocity
maps do not typically extend beyond 1-1.3 R$_{\rm e}$, the regime
reported for LTG in measuring V$_{2.2}$.  Therefore, the data for ETGs
presented in \Fig{figDM} are likely lower limits; still, the latter
probe ETG dynamics in a regime of maximal `baryons''.

\begin{figure}
\centering
\includegraphics[width=0.48\textwidth]{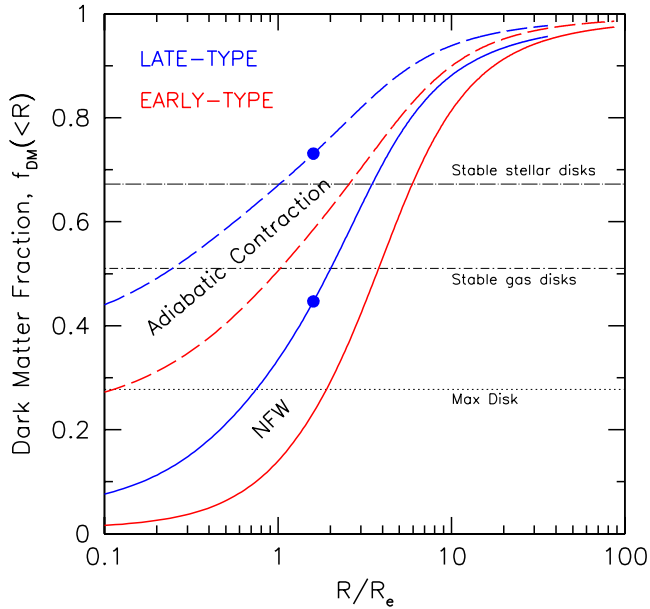}
\caption{Dark matter fraction as a function of half-light radius,
  $R_{\rm e}$, for early-type (red lines) and late-type (blue lines)
  galaxies with circular velocity $V_{\rm circ}(R_{\rm e})\simeq 230 \kms$.
  The black horizontal lines are as in Figure 1. 
  Solid lines show models with NFW haloes that follow the
  concentration-mass relation from $\Lambda$CDM
  \citep{Maccio08}. Dashed lines show models with adiabatically
  contracted NFW haloes and correspondingly lighter IMFs. The blue
  dots correspond to 2.2$\Rd$ for late-type galaxies.}
\label{fig:figDMr}
\end{figure}

\Fig{figDMr} shows dark matter fractions vs radius for $\Lambda$CDM
based models of early and late-type galaxies with circular velocity
$V_{\rm circ}(R_{\rm e})\simeq 230 \kms$ from \citet{Dutton11a}. This shows that
for a given halo response, early-type galaxies have lower dark matter
fractions than late-type galaxies. These differences are driven by
both size (early-types are smaller) and IMF (early-types have heavier
IMFs) variations.

\section{Future Outlook}

The trends observed in \Fig{figDM} and the characterization of
baryon-DM transitions in galaxies over a range of Hubble types are
significant constraints that galaxy formation models have yet to
uniquely predict (e.g. \Fig{figDMr}).  This is partly due to
additional complications such as the degeneracy between the baryon or
dark matter fractions, the stellar initial mass function, adiabatic
contraction, and various counter-acting non-gravitational agents such
as feedback and dynamical friction
\citep{Dutton11a,Trujillo-Gomez11,Oguri14}.  The advent of new
extensive dynamical models of galaxies based on wide-field integral
field surveys, such as CALIFA, MaNGA, SAMI (out to $\sim 2 R_{\rm e}$)
and SLUGGS (reaching $\sim 6-10 R_{\rm e}$) heralds a promising future
for mapping the mass distributions at all radii in galaxies. Model
dependencies and biases could at least be partially offset with reduced
sampling errors.

\section*{Acknowledgments}

\noindent
S.C. acknowledges support from the Natural Science and Engineering
Research Council of Canada through a generous Research Discovery Grant.
He is also grateful to the Instituto de Astrof\'isica de Canarias
for their hospitality while this paper was being written.
A.A.D. acknowledges support from the Sonder-forschungsbereich
SFB 881 ``The Milky Way System'' (subproject A1) of the German
Research Foundation (DFG).

\end{document}